# The reliable solution and computation time of variable parameters Logistic model

Wang Pengfei[1,2]    Pan Xinnong[3,4]

1. Center for Monsoon System Research, Institute of Atmospheric Physics (CMSR), Chinese Academy of Sciences, Beijing 100190, China

2. State Key Laboratory of Numerical Modeling for Atmospheric Sciences and Geophysical Fluid Dynamics (LASG), Institute of Atmospheric Physics, Chinese Academy of Sciences, Beijing 100029, China

3. University of Chinese Academy of Sciences, Beijing 100049, China

4. Key Laboratory for Atmosphere and Global Environment Observation (LAGEO), Institute of Atmospheric Physics, Chinese Academy of Sciences, Beijing 100029, China

Corresponding author address: Dr. Pengfei Wang, CMSR&LASG/IAP/CAS, P. O. Box 9804,

Beijing 100029, China. Email: wpf@mail.iap.ac.cn

**Abstract**

The reliable computation time (RCT, marked as Tc) when applying a double precision computation of a variable parameters logistic map (VPLM) is studied. First, using the method proposed, the reliable solutions for the logistic map are obtained. Second, for a time-dependent non-stationary parameters VPLM, 10000 samples of reliable experiments are constructed, and the mean Tc is then computed. The results indicate that for each different initial value, the Tcs of the VPLM are generally different. However, the mean Tc trends to a constant value once the sample number is large enough. The maximum, minimum and probable distribution function of Tc is also obtained, which can help us to identify the robustness of applying a nonlinear time series theory to forecasting while using the VPLM output. In addition, the Tc of the fixed parameter experiments of the logistic map was obtained, and the results suggested that this Tc matches the theoretical formula predicted value.

# 1 introduction

The logistic map was originally introduced to study biological growth in ecology, and, currently, it has become a classical chaotic system in the study of chaotic dynamics and nonlinear prediction(May 1976). In recent years, along with the development of non-stationary theory, the variable parameters logistic map (VPLM) was introduced and used to study the effects of non-stationary forcings and their predictability(Wang 2010). Studies indicate that many climate systems have non-stationary phenomena; therefore, the investigation of this simple driven forcing model is valuable for helping to understand the rules of the real climate system.

The classical logistic model is $x_{i+1} = ax_i(1-x_i)$. When this model is applied to investigations, the influence of rounding errors is often ignored. Oteo(2007) studied the influence of a double-precision floating-point computation for a fixed parameter ($a$=4) logistic map. He found that when $i$ is above 50, the result of double precision deviates from the correct result. Oteo's work suggests that we can use a quantitative experiment to study the effects of rounding errors for a discrete dynamical system. In addition, this experiment can be used to analyze the reliable computation time of the logistic map in a double precision computation environment.

Previous studies mainly focus on the fixed parameters logistic map system (i.e., $a$ is constant). However, in practical applications, we will consider a logistic map that controls parameters that are time-varying when considering the external driving force (Wang et al. 2010,2011).

$$x_{i+1} = u_i x_i (1 - x_i), \tag{1}$$

$$u_i = 3.95 - 0.4 e^{-2.5i \frac{4\pi}{25000}}, \text{ (i=1,2,…,2000)} \tag{2}$$

This type of VPLM system will also be affected by rounding errors when simulated in a double precision environment; however, generally the changing rule of this map is more complicated than that of the fixed parameters map. If we do not achieve the reliable numerical solution and know its effective computing time, we may misunderstand the conclusions deduced from the VPLM output. In this study, we apply the reliable computation method to obtain its reliable solutions, and then we obtain its mean effective computation time using 10000 samples for a double-precision float-point computation.

**2 method**

To address the lack of float-point computation accuracy, researchers have developed a software library named 'MP' (multiple precision)(Brent 1978; Oyanarte 1990). Wang et al.(Wang et al. 2006; Wang et al. 2012) applied this MP tool to investigate the Lorenz equation, and their results indicate that increasing the float-point precision is valuable to obtain the reliable solution for a chaotic system. To obtain the reliable solution for the VPLM system, we use the 2000 bits (or above) precision MP program for computation. This procedure is the same as the procedure we applied for the fixed parameters logistic map. Once we obtain the reliable solution for the VPLM, we then apply it to evaluate the reliable computation time (Tc) in a double precision computation environment.

There are two ways to obtain the reliable computation time of a chaotic system. One is the theoretical analysis, and the other is by numerical experiments. The previous studies indicate that when $a$=4.0, the Lyapunov exponent of the logistic map ($x_{i+1} = ax_i(1-x_i)$) is 0.693[9]. The analytical solution for this map (Oteo et al. 2007) is $x_n = \sin^2(2^n \arcsin\sqrt{x_0})$, which can be regarded as the cross-point of the continuous functions $x(t) = \sin^2(2^t \arcsin\sqrt{x_0})$ and $y(t)=t$ for positive integer numbers. The subscript $i$ in function $x$ can be regarded as time (unit: second). Once we know the above relations, we can apply the theoretical formula between the reliable computation time $T_c$ and the Lyapunov exponents to compute the $T_c$ of the logistic map. Wang, etc. (Wang et al. 2014) show that the relationship of $T_c$ and the Lyapunov exponents is $T_c \approx \frac{\ln B}{\lambda} K$, where $B$ is the base number of the float-point computation. For example, $B$=2 is binary, $B$=10 is decimal and $K$ is the number of significant digits in each system. For a double precision computation in which $B$=2 and $K$=53, after substituting these values into the formula, we obtain $T_c \approx \frac{\ln 2}{0.693} 53 \approx 53$, which indicates that the maximum effective computation time of the logistic map is approximately 53 seconds in a double precision computation. We will validate its correctness in an experiment in section 3.

Different from the fixed parameters logistic map, the difficulty in calculating $T_c$ in the VPLM system is that the Lyapunov exponents of this system vary with time; therefore, we cannot obtain their values from the theoretical formula, but we can obtain them from numerical experiments. For the result to have statistical meaning, we ran 10000 experiments as individual samples. Because the solution range of the VPLM is within [0, 1], we can obtain 10000 initial values, which are uniformly distributed in this interval. For each initial value, we compute the solution by a double precision (DP) and multi-precision (MP) computation simultaneously. The difference between the DP and MP solutions begins at zero and increases over time. When the error reaches the tolerance $\delta = 0.1$, we denote this time as the reliable computation time, corresponding to this initial value. Finally, we obtain 10000 Tc values for 10000 samples and then can obtain the mean reliable computation time.

**3 result**

We use the initial value $x_0 = 0.1$ to compute the logistic map

$$x_{i+1} = 4x_i(1-x_i). \tag{3}$$

First, we apply a double precision computation to obtain a series of $x$ ($i$= 1, 2,..., 2000) and then, using the MP program, compute the corresponding reliable solution as the reference solution. In this experiment, the computation precision is set as 4096, which is sufficient to theoretically obtain the first 2000 reliable data (in Table 1, we only listed the first 100 values. All of our programs are written in C language and run on a Linux system. The DP solution was computed by C language internal functions, and the MP solution was computed by the MP library).

As shown in table 1, when $i = 51,52,53$, the solution of the DP computation is significantly different than the MP solution. This fact indicates that after that moment, the double precision solution is no longer reliable, and, thus, Tc can be regarded as 53. This value matches the theoretical result well. In Figure 1b, we plotted an instance of the first 100 steps for DP and MP. The blue curve represents the DP computation, and the red curve represents the MP computation. The absolute error between DP and MP is shown in figure 1a, and it continues to increase from $10^{-15}$ to $10^{-1}$ and then oscillates at approximately $10^{-1}$.

When *a* is a constant smaller than 4.0, the Lyapunov exponent of the logistic map is generally smaller than 0.693. However, as long as the system is in the chaotic region, the maximum Lyapunov exponent is generally a positive constant. Thus, we can address these cases with a similar procedure as for *a*=4.

The methods referenced in section 2 are applied to obtain the reliable computing time of the non-stationary time series outputted by the VPLM (equation 1-2). The solutions for the two different initial values (from 10000 initial values) are listed in table 2 and 3. From these results, we found that by using different initial values, the Tc values of the VPLM are different as well. Thus, to obtain a statistically meaningful result, we should use sufficient samples to calculate the mean $T_c$. In figure 2, we have plotted the first 300 steps of the DP and MP computations for the VPLM (see FIG. 2b and 2d for the different initial values). The absolute error between the DP and MP computation is shown in figure 2a and 2c. The error reaches $10^{-15}$ after 50 steps and then increases to $10^{-1}$ slowly. In addition, the error curve in figure 2a and 2c are not exactly the same, which means that they have different Tc values.

The PDF distribution of Tc for 10000 samples is shown in figure 3, in which most of the Tc values are located in [190-270]. The maximum Tc is 303, the minimum Tc is 168 and the statistical mean of Tc is 231.66 (the author used $10^5$ samples to compute the mean Tc, and the result is 231.10, which is almost same as that of the $10^4$ samples. Therefore, we can conclude that 10000 samples are sufficient). The mean Tc (231.66) indicates that when the computation time is beyond 231.66, the double precision solution is no longer credible. In other words, the 231 (at most 303) continuous steps in the solution series follow the VPLM rules, and once the data is beyond, these steps tend to become unrelated (although we know they are not accurate, we still cannot determine the error at each step, and the only way to estimate the error is using a numerical experiment). Therefore, using these 2000 double precision solutions as time series to study nonlinear prediction problems will cause the data quality to be questionable.

In summary, for any initial value of the VPLM map, we can obtain two series of data by the DP and MP method. One series is a not very accurate double precision solution (accurate data is obtained only when using continuous 231 steps, and data is generally accurate beyond 303 steps as shown in figure 3), and another is a reliable series for all 2000 steps. We hope researchers apply this reliable series of the VPLM to further research and make the results more reliable and convincing.

# Acknowledgements

This work is supported by the National Natural Sciences Foundation of China (41375112 and 41530426) and the CAS Key Technology Talent Program.

*float-point precision and the Lyapunov exponent in chaotic systems"*. arXiv preprint arXiv:1410.4919.

**Table 1. The first 100 steps of Logistic map for $x_0 = 0.1$ and a=4.**

| $i$ | DP | MP | $i$ | DP | MP |
| --- | --- | --- | --- | --- | --- |
| 0 | 0.100000000000000 | 0.100000000000000 | 50 | 0.560036763222377 | 0.539205854891865 |
| 1 | 0.360000000000000 | 0.360000000000000 | 51 | 0.985582348247121 | 0.993851603768792 |
| 2 | 0.921600000000000 | 0.921600000000000 | 52 | 0.056839132283247 | 0.024442373819968 |
| 3 | 0.289013760000000 | 0.289013760000000 | 53 | 0.214433781298139 | 0.095379776728053 |
| 4 | 0.821939226122650 | 0.821939226122650 | 54 | 0.673807738945284 | 0.345129899677439 |
| 5 | 0.585420538734197 | 0.585420538734198 | 55 | 0.879163479530912 | 0.904061008104319 |
| 6 | 0.970813326249438 | 0.970813326249438 | 56 | 0.424940223160047 | 0.346938806918886 |
| 7 | 0.113339247303761 | 0.113339247303763 | 57 | 0.977464119602946 | 0.906289084690344 |
| 8 | 0.401973849297512 | 0.401973849297517 | 58 | 0.088112057967135 | 0.339716718645931 |
| 9 | 0.961563495113813 | 0.961563495113816 | 59 | 0.321393292831724 | 0.897237078873489 |
| 10 | 0.147836559913285 | 0.147836559913272 | 60 | 0.872398576618023 | 0.368810812672229 |
| 11 | 0.503923645865164 | 0.503923645865127 | 61 | 0.445277200531482 | 0.931157588513116 |
| 12 | 0.999938420012499 | 0.999938420012500 | 62 | 0.988021660873313 | 0.256412535470218 |
| 13 | 0.000246304781624 | 0.000246304781619 | 63 | 0.047339434073811 | 0.762660588495809 |
| 14 | 0.000984976462315 | 0.000984976462296 | 64 | 0.180393648221529 | 0.724037661004141 |
| 15 | 0.003936025134734 | 0.003936025134657 | 65 | 0.591407119611426 | 0.799228505807174 |
| 16 | 0.015682131363489 | 0.015682131363186 | 66 | 0.966578953937370 | 0.641849205249625 |
| 17 | 0.061744808477550 | 0.061744808476377 | 67 | 0.129216318970838 | 0.919515211880199 |
| 18 | 0.231729548414484 | 0.231729548410370 | 68 | 0.450077847529859 | 0.296027948004448 |
| 19 | 0.712123859224412 | 0.712123859215584 | 69 | 0.990031114770992 | 0.833581608018896 |
| 20 | 0.820013873390967 | 0.820013873405948 | 70 | 0.039478026225196 | 0.554893243166111 |
| 21 | 0.590364483349242 | 0.590364483310888 | 71 | 0.151678046682236 | 0.987946927418825 |
| 22 | 0.967337040596098 | 0.967337040623825 | 72 | 0.514687267347589 | 0.047631184090112 |
| 23 | 0.126384361947522 | 0.126384361843861 | 73 | 0.999137136711442 | 0.181449817569144 |
| 24 | 0.441645420010560 | 0.441645419700724 | 74 | 0.003448475022014 | 0.594103125093074 |
| 25 | 0.986378971977024 | 0.986378971832381 | 75 | 0.013746332168146 | 0.964578407390869 |
| 26 | 0.053741982474292 | 0.053741983037102 | 76 | 0.054229482080276 | 0.136667613544656 |
| 27 | 0.203415127176100 | 0.203415129185368 | 77 | 0.205154581414323 | 0.471958307810658 |
| 28 | 0.648149652848124 | 0.648149657615473 | 78 | 0.652264716556147 | 0.996854653996633 |
| 29 | 0.912206721443921 | 0.912206715793673 | 79 | 0.907261824368305 | 0.012541811207546 |
| 30 | 0.320342475185814 | 0.320342493818377 | 80 | 0.336551225648800 | 0.049538056716719 |
| 31 | 0.870892695110561 | 0.870892721890401 | 81 | 0.893137992652362 | 0.188336150613801 |
| 32 | 0.449754434854499 | 0.449754355394920 | 82 | 0.381770074933085 | 0.611462579943107 |
| 33 | 0.989901532732837 | 0.989901500792880 | 83 | 0.944086739274687 | 0.950304373089706 |
| 34 | 0.039985952904069 | 0.039986078083536 | 84 | 0.211147872001506 | 0.188903886305149 |
| 35 | 0.153548305897691 | 0.153548766572132 | 85 | 0.666257792602967 | 0.612876832175841 |
| 36 | 0.519884894614559 | 0.519886171425237 | 86 | 0.889433385595155 | 0.949035283031788 |
| 37 | 0.998418363864672 | 0.998418160744184 | 87 | 0.393366552735581 | 0.193469258370249 |

| 38 | 0.006316538249856 | 0.006317348161538 | 88 | 0.954517231698025 | 0.624155617743659 |
| --- | --- | --- | --- | --- | --- |
| 39 | 0.025106558377575 | 0.025109757094974 | 89 | 0.173656344358254 | 0.938341530330762 |
| 40 | 0.097904876416033 | 0.097917028774422 | 90 | 0.573999273689525 | 0.231426811149144 |
| 41 | 0.353278046359976 | 0.353317137001643 | 91 | 0.978096429973691 | 0.711473768921930 |
| 42 | 0.913890673280218 | 0.913936550810420 | 92 | 0.085695214585646 | 0.821115380231816 |
| 43 | 0.314778042286590 | 0.314626127612689 | 93 | 0.313406179131065 | 0.587539650314305 |
| 44 | 0.862771305523247 | 0.862546109744532 | 94 | 0.860730984054127 | 0.969347238491397 |
| 45 | 0.473587919555836 | 0.474241273236424 | 95 | 0.479492628573366 | 0.118852678882000 |
| 46 | 0.997209608026444 | 0.997345951982078 | 96 | 0.998317790868680 | 0.418906878418289 |
| 47 | 0.011130422744760 | 0.010588016188163 | 97 | 0.006717517215034 | 0.973695622528535 |
| 48 | 0.044026145737131 | 0.041903640405449 | 98 | 0.026689568709997 | 0.102449828789213 |
| 49 | 0.168351376914654 | 0.160590901304880 | 99 | 0.103908942528286 | 0.367815445481095 |
|  |  |  | 100 | 0.372447496763758 | 0.930108974186555 |

**Table 2. The first 300 steps (interval 5) for VPLM with $x_0 = 0.1$.**

| $i$ | DP | MP | $i$ | DP | MP |
|---|---|---|---|---|---|
| 0 | 0.100000000000000 | 0.100000000000000 | 155 | 0.857821626017957 | 0.857821909330965 |
| 5 | 0.832592851312631 | 0.832592851312630 | 160 | 0.535112449500775 | 0.535110168030664 |
| 10 | 0.543381756879222 | 0.543381756879222 | 165 | 0.871584758242825 | 0.871581509329935 |
| 15 | 0.889167004162284 | 0.889167004162284 | 170 | 0.411382702016328 | 0.411414567709174 |
| 20 | 0.371262401351000 | 0.371262401351000 | 175 | 0.891043235648514 | 0.891101146029621 |
| 25 | 0.808210315973505 | 0.808210315973505 | 180 | 0.307555675733818 | 0.307706004445068 |
| 30 | 0.493520322668974 | 0.493520322668974 | 185 | 0.907610248538264 | 0.907516514654172 |
| 35 | 0.880720968984047 | 0.880720968984047 | 190 | 0.505630697151845 | 0.503674463640026 |
| 40 | 0.344514552359543 | 0.344514552359542 | 195 | 0.825715927618323 | 0.825110829796234 |
| 45 | 0.837428219346418 | 0.837428219346419 | 200 | 0.639706618737391 | 0.637623080264878 |
| 50 | 0.560532314789229 | 0.560532314789228 | 205 | 0.763039767716258 | 0.764730506671172 |
| 55 | 0.893029925860431 | 0.893029925860432 | 210 | 0.313848497119786 | 0.307251305171718 |
| 60 | 0.376294800158017 | 0.376294800158021 | 215 | 0.895113865546431 | 0.909230443671585 |
| 65 | 0.803664822317961 | 0.803664822317964 | 220 | 0.309079653805944 | 0.521664114388859 |
| 70 | 0.476822440123461 | 0.476822440123481 | 225 | 0.906406917714885 | 0.821930005527828 |
| 75 | 0.881249771116657 | 0.881249771116646 | 230 | 0.455241622719538 | 0.639690235714764 |
| 80 | 0.335983844868298 | 0.335983844868324 | 235 | 0.863848044506658 | 0.755689159659303 |
| 85 | 0.849305654962446 | 0.849305654962366 | 240 | 0.531067914806285 | 0.357202934866837 |
| 90 | 0.545717901170099 | 0.545717901170505 | 245 | 0.826904347375071 | 0.750558962175503 |
| 95 | 0.893869699914088 | 0.893869699914330 | 250 | 0.671285749390229 | 0.407925911347316 |
| 100 | 0.360317549873014 | 0.360317549875023 | 255 | 0.818861455246060 | 0.903010288758529 |
| 105 | 0.791035400810154 | 0.791035400808953 | 260 | 0.636716065663608 | 0.374800835100522 |
| 110 | 0.398014367631752 | 0.398014367624153 | 265 | 0.754635040717098 | 0.790407952675023 |
| 115 | 0.847201585829368 | 0.847201585810587 | 270 | 0.364305878836527 | 0.381871892174048 |
| 120 | 0.575362825035316 | 0.575362825120107 | 275 | 0.755626304670800 | 0.820284054110586 |
| 125 | 0.897456100113565 | 0.897456100042480 | 280 | 0.354012774381064 | 0.654803888170862 |
| 130 | 0.378055022881996 | 0.378055022066093 | 285 | 0.737419723296179 | 0.743968277468713 |
| 135 | 0.799985544913794 | 0.799985543226208 | 290 | 0.588714334137301 | 0.494287176942652 |
| 140 | 0.461259918019212 | 0.461259905376383 | 295 | 0.915665309466767 | 0.748056014753772 |
| 145 | 0.881745862476446 | 0.881745877870685 | 300 | 0.632922589329766 | 0.439826720591001 |
| 150 | 0.327704384756794 | 0.327704309112235 | | | |

**Table 3. The first 300 steps (interval 5) for VPLM with $x_0 = 0.6$.**

| $i$ | DP | MP | $i$ | DP | MP |
|---|---|---|---|---|---|
| 0 | 0.600000000000000 | 0.600000000000000 | 155 | 0.379315744903887 | 0.379315177127178 |
| 5 | 0.380275886117672 | 0.380275886117672 | 160 | 0.802330634477705 | 0.802329191782283 |
| 10 | 0.813747928284932 | 0.813747928284932 | 165 | 0.482557000231579 | 0.482545591306503 |
| 15 | 0.513031780808352 | 0.513031780808351 | 170 | 0.847606678155340 | 0.847617473433552 |
| 20 | 0.882776067652321 | 0.882776067652321 | 175 | 0.611106990260610 | 0.611056879154665 |
| 25 | 0.348109006292028 | 0.348109006292028 | 180 | 0.831884841964112 | 0.832043962837539 |
| 30 | 0.833516133790997 | 0.833516133790998 | 185 | 0.644851604897976 | 0.645025950622436 |
| 35 | 0.556298328300335 | 0.556298328300334 | 190 | 0.767487963778570 | 0.767597929156230 |
| 40 | 0.891744714450323 | 0.891744714450324 | 195 | 0.303679969208342 | 0.303485506961593 |
| 45 | 0.374724087810298 | 0.374724087810300 | 200 | 0.909342022583631 | 0.909245300737688 |
| 50 | 0.804636682562754 | 0.804636682562755 | 205 | 0.533468011769676 | 0.531327332478798 |
| 55 | 0.479964749321297 | 0.479964749321301 | 210 | 0.850981111722547 | 0.846999365380964 |
| 60 | 0.881696083660928 | 0.881696083660927 | 215 | 0.621633988051341 | 0.641980708338287 |
| 65 | 0.337818000131651 | 0.337818000131652 | 220 | 0.801059396282309 | 0.757899798173127 |
| 70 | 0.848068286414706 | 0.848068286414703 | 225 | 0.480425307518706 | 0.341301839310297 |
| 75 | 0.542381878128891 | 0.542381878128908 | 230 | 0.815548182091915 | 0.777559693054478 |
| 80 | 0.892240765947636 | 0.892240765947644 | 235 | 0.604003887225542 | 0.302849243722797 |
| 85 | 0.356087782990957 | 0.356087782991019 | 240 | 0.867639715592924 | 0.913439234124305 |
| 90 | 0.799789651568964 | 0.799789651568882 | 245 | 0.491764355139892 | 0.605781247588618 |
| 95 | 0.455542260881015 | 0.455542260880479 | 250 | 0.788430125316397 | 0.863983545725774 |
| 100 | 0.892918171866280 | 0.892918171866642 | 255 | 0.366033267615156 | 0.539483465449693 |
| 105 | 0.350227999722879 | 0.350227999725469 | 260 | 0.761916085661156 | 0.838507709569241 |
| 110 | 0.802646029881655 | 0.802646029876441 | 265 | 0.308511986122180 | 0.697376860807326 |
| 115 | 0.477238556590487 | 0.477238556554562 | 270 | 0.905464647051084 | 0.915729024268197 |
| 120 | 0.871430513957365 | 0.871430513986760 | 275 | 0.408535223266365 | 0.646963784101408 |
| 125 | 0.397165700828442 | 0.397165700580933 | 280 | 0.909395308242515 | 0.738716952924999 |
| 130 | 0.847663050479794 | 0.847663049834265 | 285 | 0.485242724850372 | 0.568578491212588 |
| 135 | 0.583308290087215 | 0.583308293106579 | 290 | 0.761749450197490 | 0.903377890996209 |
| 140 | 0.891620997474507 | 0.891620993130455 | 295 | 0.304159596724964 | 0.360082686274566 |
| 145 | 0.323367391000731 | 0.323367371492267 | 300 | 0.914437554171898 | 0.741677426184428 |
| 150 | 0.874508295910031 | 0.874508361230068 | | | |

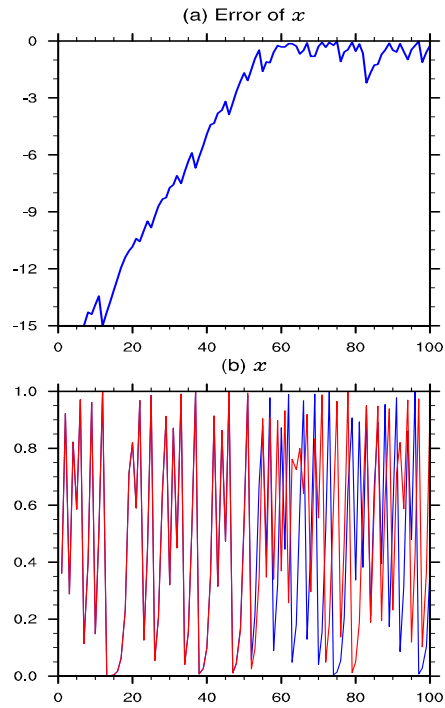

Figure 1. Numerical solution of the fixed parameters logistic map. The initial value is 0.1. (a) The logarithm of the absolute error between DP and MP for the first 100 steps ($\log_{10}$); (b) The solution of DP and MP for the first 100 steps; the blue and red curves represent DP and MP, respectively.

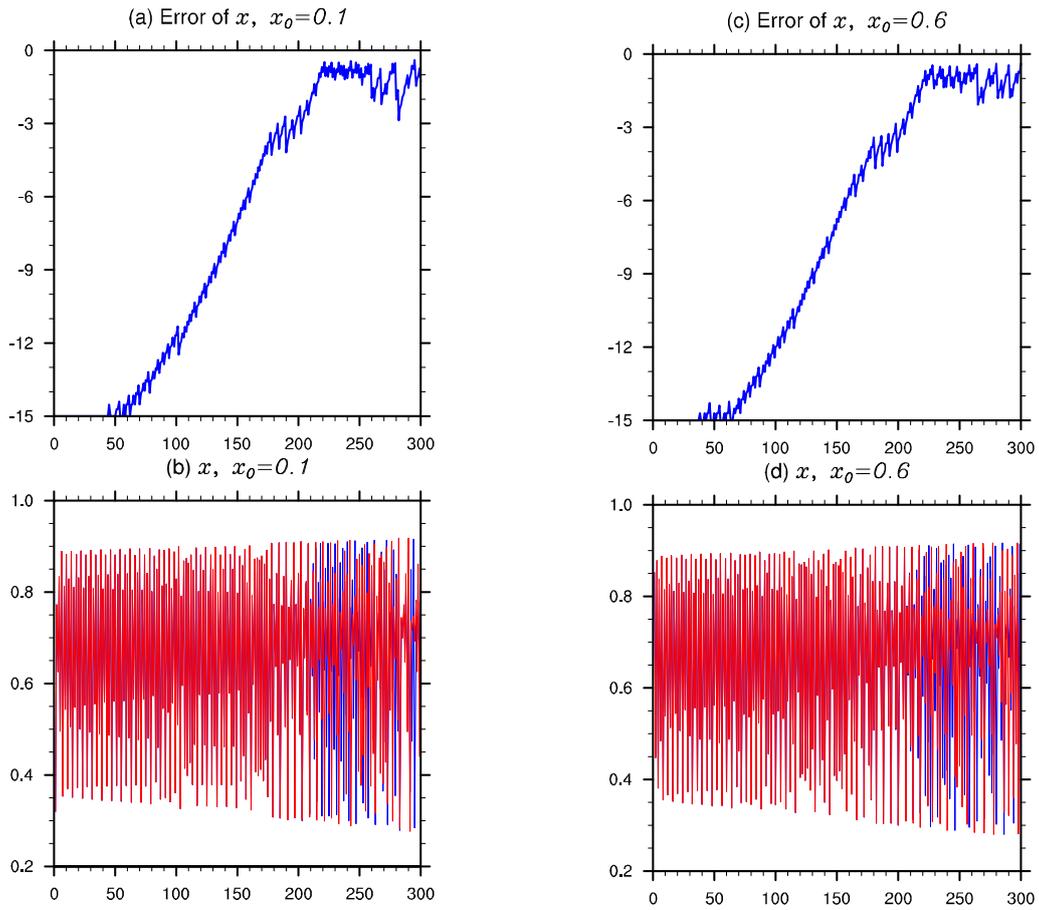

Figure 2. Numerical solution of VPLM; the initial value of (a, b) is 0.1. (a) The logarithm of the absolute error between the DP and MP computations for the first 300 steps ($\log_{10}$), (b) the solution of the DP and MP computations for the first 300 steps; blue and red curves represent the DP and MP computations, respectively; (c, d) is same as (a, b), but for the initial value setting of 0.6.

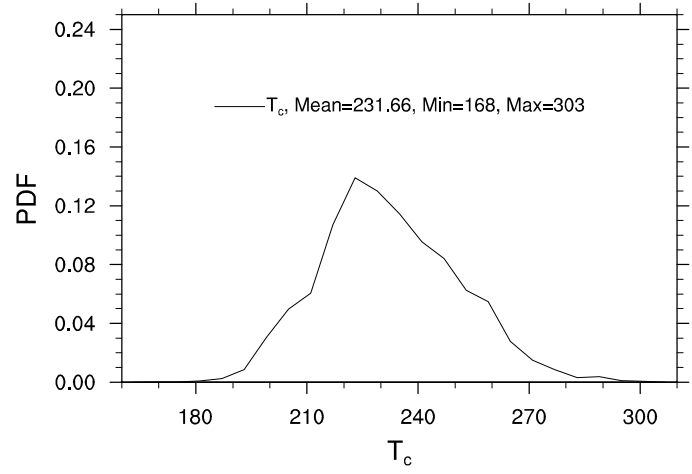

Figure 3. The PDF distribution of the reliable computation time for the VPLM.